\documentstyle[12pt]{article}
\newtheorem{thm}{Theorem}

\newtheorem{lem}{Lemma}
\newtheorem{df}{Definition}

 \newcommand{\Integer}{\:\mbox{\sf Z} \hspace{-0.82em} \mbox{\sf Z}\,}
  
 \newcommand{\Real}{\mbox{I \hspace{-0.82em} R}}
 
 \newcommand{\Complex}{
        \mbox{C \hspace{-1.16em} \raisebox{-0.018em}{\sf l}}\;}
 \newcommand{\Z}{\Integer}

 \newcommand{\bd}{\begin{df}}
 \newcommand{\bt}{\begin{thm}}
 \newcommand{\bl}{\begin{lem}}
 \newcommand{\ed}{\end{df}}
 \newcommand{\be}{\begin{equation}}
 \newcommand{\ee}{\end{equation}}
 \newcommand{\bi}{\begin{itemize}}
 \newcommand{\ei}{\end{itemize}}
 \newcommand{\et}{\end{thm}}
 \newcommand{\el}{\end{lem}}
 \newcommand{\bs}{\bigskip}
 \newcommand{\ms}{\medskip}
 \newcommand{\eb}{\hfill$\Box$}
 \newcommand{\tl}{{\cal T}}
 
 \newcommand{\tr}{\hbox{tr}}
 \newcommand{\pkt}{puncture}
 \newcommand{\K}{{\cal K}}

 \newcommand{\A}{{\cal A}}
 \newcommand{\2}{^{pct}}
 
 \newcommand{\mi}{pattern}
 \newcommand{\mii}{oriented pattern class}
 
 \newcommand{\ti}{tile}

 \newcommand{\B}{{\cal B}}

 \newcommand{\Om}{\Omega}

 \newcommand{\Gr}{{\cal R}}
 
 \newcommand{\So}{Schr\"odinger operator}

 \newcommand{\sst}{substitution}

 \newcommand{\CA}{$C^*$-algebra}

\newcommand{\bet}{\check{\beta}}

\newcommand{\si}{stably isomorphic}
\newcommand{\mk}{decorat}
\newcommand{\pe}[1]{$#1$-facet}
\newcommand{\faden}{rope}

 \title{
 \bf Gap Labelling for Schr\"odinger Operators
 on Quasiperiodic Tilings
   \vspace{1.5em}}
 \author{Johannes Kellendonk}
 \date{Department of Mathematics, King's College London,\\
       Strand, London WC2R 2LS\\
        \vspace{.2em}
       {\small E-mail: johannes@mth.kcl.ac.uk}}

\begin{document}
 \maketitle

 \begin{abstract}
\noindent
For a large class of tilings, including those which are obtained by the
 generalized dual method from regular grids, it is shown
that their algebra is stably isomorphic to a crossed product with $\Z^d$.
For $d\leq 3$ this in particular enlarges the class of tilings
of which can be shown
that a set of possible gap labels is completely determined by
an invariant measure on the hull.

 \end{abstract}
 \begin{flushright}
 \parbox{12em}
  { \begin{center}
      KCL-TH-94-10
 \end{center} }
 \end{flushright}

\newpage

\bibliographystyle{unsrt}

\section*{Introduction}

\addcontentsline{toc}{section}{\bf Introduction}

Quasiperiodic tilings\footnote{
We simply call the analogs of tilings in other dimensions than two
(e.g.\ sequences and packings)
tilings, too.} serve as models for the spatial structure of quasicrystals
even though they are highly idealized.
In particular, tilings with non trivial orientational order
are of interest and the so-called generalized dual method (GDM) is a powerful
way to create such tilings among which the Penrose tilings are the most
famous ones.

The gap labelling of discrete \So s on tilings in its $C^*$-algebraic
formulation
\cite{Be1} is a step towards an
understanding of the nature of the spectrum of such operators as well
as of the influence of the geometry of the tiling on them.
This is of particular interest
for operators on higher dimensional
non-periodic tilings for which a computation of their
spectra is presently out of reach.
In this formulation, where the  scaled ordered $K_0$-group of a
\CA\ with a normalized trace is used,
the gap labelling does not refer to a specific operator but only involves
this algebra. The natural candidate for this algebra
is the algebra
$\A_\tl$ of the tiling $\tl$ on which the \So\ is defined. It
contains all local operators involving translations and multiplications
with pattern dependent functions.
A local selfadjoint operator $H$ on the tiling $\tl$ is then an operator in
a specific representation of $\A_\tl$, namely on the space of
(square summable) wavefunctions $\psi$ taking values on the \ti s of $\tl$
\be \label{14061}
H\psi(x)=\sum_{x'} H_{x,x'}\psi(x').
\ee
Here
$x$, $x'$ denote \ti s of $\tl$ and locality refers to the requirement that
the matrix element $H_{x,x'}$ depends only on a the \mii \
of a certain radius to which $x,x'$ belong.
The values of its integrated density of states (IDS) $N_H(E)$ at energies
$E$ lying in a gap of its spectrum serve as labels for the gaps;
they are insensitive to certain perturbations of the operator.
The abstract gap labelling of Bellissard \cite{Be1} states
that, if $E$ lies in a gap, then
\be
N_H(E)\in \tr_*K_0(\A_\tl)\cap[0,1],
\ee
where $K_0(\A_\tl)$ is the $K_0$-group of
$\A_\tl$ and $\tr_*$ the state on it induced by the trace $\tr$.
It requires the validity of Shubin's formula by which this trace is
equated to the operator trace per volume in that representation.
One  motivation for using $\A_\tl$ is that
it is expected not to yield to many values on the r.h.s.\ so that for
``generic'' $H$
all elements of $\tr_*K_0(\A_\tl)$ actually occur as labels for
 open gaps of $H$. In that case,
and if $\tr$ is faithful on $\A_\tl$,
the density of the values of the IDS on gaps in $[0,1]$
expresses the fact that the continuous part of the
spectrum is a Cantor set.
\ms

The main result of this work is the
extension of the equality
\be         \label{08061}
\tr_*K_0(\A_\tl)=\mu(C(\Om,\Z)),
\ee
$\mu$ being the invariant measure on the hull of the tiling corresponding to
the trace, to a large class of tilings including those which are obtained
by the GDM from regular grids. In particular this fills a gap in \cite{Ke2}
for the gap labelling of the Penrose tilings where (\ref{08061}) has only been
conjectured.

The explicit computation of such measures, which are in known cases probability
measures for the frequencies of patterns occurring in the tilings, is not
addressed here. Let us only mention that these may be successfully
computed if the tiling allows for a \sst\ \cite{Q,BBG,Ke2}.\ms

Apart from the application to the gap labelling the computation of
the $K$-groups of $\A_\tl$
including order and scale of $K_0$
is of interest for the characterization of the tiling:
they are the topological invariants in the non commutative geometrical
description of the tiling \cite{Con}.\bs

In the first section the construction of $\A_\tl$ is reviewed taking a slightly
different point of view from \cite{Ke2}
in the description of the groupoid.
In the second, tilings that are decorations of $\Z^d$ are looked at. Their
algebra is a crossed product with $\Z^d$. It is for a subclass of these tilings
that (\ref{08061}) could be rigorously proven so far. In the third section
the concept of a reduction of a tiling is introduced. It is designed to obtain
a \si\ algebra which therefore has the same $K$-groups and the same order.
This  is used to extend the validity of (\ref{08061}),
and in section four many examples for this are given by showing
that any tiling obtained by the GDM from a regular grid has a reduction
which is a decoration of $\Z^d$.

\section{The tiling algebra $\A_\tl$}

As the algebra of a tiling is a groupoid-\CA\ groupoids play a central
role. The groupoids, which are defined by the spatial structure of the
tilings, are given by equivalence relations, i.e.\
its elements are pairs of equivalent elements of some set $X$, multiplication,
which is only defined for pairs $(x,y),(x',y')$ if $x'=y$,
is given by $(x,y)(y,z)=(x,z)$, and inversion by
$(x,y)=(y,x)^{-1}$.
In the terminology of \cite{Ren} these are all principal groupoids.
The topology of the groupoids in question needs in
general not to coincide with the relative topology from $X\times X$.

Sometimes the equivalence may be expressed as orbit equivalence under
the (right) action $\varphi$
of a group $S$ acting on $X$: $x\sim y$ whenever
$\exists s\in S:y=\varphi(s)x$.
This leads to the consideration of another kind of groupoid which is called
transformation group in \cite{Ren}.
Its space is the Cartesian product of $X$ with $S$ (here always considered to
carry the product topology) and the groupoid
structure is defined by $(x,s)(x',t)=(x,st)$ provided $x'=\varphi(s)x$ and
$(x,s)^{-1}=(\varphi(s)x,s^{-1})$.
We write it as $X\times_\varphi S$. It
may be viewed as the groupoid defined by orbit equivalence
only if $S$ acts freely on $X$.
\bs

Let us review the construction of the groupoid of a tiling $\tl$ from a
slightly more abstract point of view as in \cite{Ke2}.
A $d$ dimensional tiling
$\tl$ of $\Real^d$ is a complete covering of $\Real^d$ by \ti s
which do not overlap (more precisely the only overlap between neighboured
tiles lies in their boundary).
These \ti s, which for simplicity may
be seen as polyhedra, may have an additional decoration, e.g.\ to break
symmetries. However it is not the precise form of the \ti s which is important
but their spatial arrangement. A tiling encodes a set of allowed translations
in $\Real^d$ as being those which lead from one \ti\ to another. They do in
general not form a group. In order to get a hold on the allowed translations,
a point in each \ti\ of $\tl$ was identified  in \cite{Ke2}
with a point of $\Real^d$,
i.e.\ any \ti\ got a \pkt\ and translation from the \ti\ with \pkt\
$x$ to the \ti\ with \pkt\ $y$ became translation of $\tl$ in $\Real^d$
by $y\!-\!x$ the translate being denoted by $\tl-(y\!-\!x)$.
But it is quite fruitful to also consider the case in which
only part of the \ti s carry a \pkt\, thus restricting the set of allowed
translations, where it is still assumed that the \ti\ which lies on the origin
$0\in\Real^d$ is \pkt d, and its \pkt\ is denoted by $0$, too.
Next to the actual \ti s of $\tl$ and its \mi s, which are finite collections
of \ti s, we have to consider their \mii es, i.e.\ their equivalence classes
under translations. As in \cite{Ke2} we require the compactness condition
(there called \B1) to
hold, namely that there are for any given
finite size only finitely many \mii es  not exceeding that size.

The hull $\Om_\tl$ of $\tl$ is given by the closure of the
set of translates of $\tl$ by translations from the \ti\ on $0$ to any other
\pkt d \ti\ of $\tl$. Hence in particular is contains elements like $\tl-x$
which is $\tl$ translated such that the \ti\ with \pkt\ $x$ lies on $0$.
The closure is taken with respect to a metric which assigns
to two tilings a small distance if they coincide on a large $r$-ball around
$0$: let $M_r(T)$ be the \mii\ of the smallest pattern of $T$ which covers the
$r$-ball around $0\in\Real^d$ then $d(T,T')=\exp(-\sup\{r|M_r(T)=M_r(T')\})$
is the metric and
\be \label{21051}
\Om_\tl := \overline{\{\tl-x|x\in\tl\2\}}.
\ee
$\Om_\tl$ contains all those tilings which look on finite patches like
some patch of $\tl$, i.e.\ which are locally homomorphic to $\tl$.
The open neighbourhoods of an element $T$ are of the form
\be U_{M_r(T)} = \{T'\in\Om_\tl|M_r(T')=M_r(T)\} \ee
and are as smaller as larger $r$ is.
They are special kinds  of the following type of sets:
$U_{M,x}:=\{T\in \Om_\tl|(M,x)\subset (T,0)\}$,
where $(M,x)\subset (T,y)$ means that a pattern of the \mii\ $M$ occurs in
$T$ such that \pkt\ $x\in M\2$ coincides with $y\in T\2$.
All these sets
are both, open and closed, so that $\Om_\tl$, which is simply denoted by
$\Om$ if no confusion arises, is a totally disconnected compact space.
Denote by $M(x,y)$, $x,y\in\tl\2$ the \mii\ of $\tl$
which covers the smallest ball that contains $x$ and $y$. To simplify the
notation denote the \pkt\ of $M(x,y)$ which corresponds to $x\in\tl\2$ in the
above identification by $x$, too.

The set
$\Gamma = \tl\2\times\tl\2$
with discrete topology
may be seen as a
groupoid (in a rather simple way) defined by the
equivalence relation on $\tl\2$ by which all elements are equivalent.
It "acts" on $\Om$ in the
following sense. On
\be
\Om^\Gamma=\{(T,(x,y))|(M(x,y),x)\subset (T,0)\}\subset \Om\times\Gamma
\ee
with relative topology define the
map $\gamma:\Om^\Gamma\rightarrow \Om$
\be
\gamma(T,(x,y)) = T-(y\!-\!x);
\ee
in fact $(M(x,y),x)\subset (T,0)$ implies that there is a unique $y'\in T\2$
which coincides with $y\in M(x,y)\2$ in the above identification and
its corresponding translation is denoted
simply by $y\!-\!x$ as it would be if \pkt s are identified with points in
$\Real^d$.
 Clearly
$\gamma(\cdot,(x,y)):U_{M(x,y),x}\rightarrow
U_{M(x,y),y}$ is continuous.\footnote{
$\gamma$ has properties analogous to a group action, namely
$\forall T\in U_{M(x,x'),x}\cap
U_{M(x,x''),x}:\gamma(\gamma(T,(x,x')),(x',x''))=\gamma(T,(x,x''))$
and a similar expression holds for the inversion.}
The groupoid $\Gr$ assigned to the tiling $\tl$ is as a topological space
the image of $\Om^\Gamma$ under $(\pi_1,\gamma)$,
$\pi_1$ being projection onto the first factor,
\be
\Gr := (\pi_1,\gamma)(\Om^\Gamma)\subset\Om\times\Om
\ee
with weak topology induced by $(\pi_1,\gamma)$ and its groupoid structure
is defined by the equivalence relation $T\sim T'$ whenever $\exists x\in T\2:
T'=T-x$.\ms

To make contact to \cite{Ke2} let for a \mi\ $M$ of $\tl$ and $x,y\in M\2$
\be
U_{M,x,y}:=\{(T,T-(y-x))|(M,x)\subset (T,0)\}.
\ee
\bl
$\Gr$ is r-discrete and its topology is generated by the sets
$U_{M,x,y}$.
\el
{\em Proof:}
The topology of $\Om^\Gamma$ is generated by the sets
$U_{M,x}\times(x',y')$ where $(M(x',y'),x')\subset (M,x)$.
There is a unique $y\in M\2$ which coincides with $y'$ under
the identification of $M(x',y')$ with a subpattern of $M$ as above.
For this $y$ we have
$(\pi_1,\gamma)U_{M,x}\times(x',y')=U_{M,x,y}$ showing that the
topology of $\Gr$ is indeed generated as stated. Clearly $\Om=\cup_M U_{M,x,x}$
is open so that $\Gr$ is r-discrete. \eb\bs

The algebra of the tiling $\tl$ is the reduced groupoid-\CA\ $C^*_{red}(\Gr)$.
It is the closure of the $*$-algebra of continuous
functions $f:\Gr\rightarrow \Complex$
with compact support, multiplication and involution being given by
\begin{eqnarray}                             \label{25061}
f*g\,(T,T') & = & \sum_{T'' \sim T} f(T,T'')\: g(T'',T') , \\
f^*\,(T,T') & = & \overline{f(T',T)}        \label{25062}.
\end{eqnarray}
The closure is taken with respect to the norm
 $\|f\|_{red}=\sup_{T\in\Om_\tl}\|\pi_T(f)\|$
the supremum being taken over all
representations $\pi_T$, $T\in\Om_\tl$, of the form
\be                                           \label{225}
(\pi_T(f)\psi)(T') := \sum_{T''\sim T'} f(T',T'')\psi(T'') ,
\ee
where $\pi_T$ acts on (square summable) wave functions
$\psi:\{T-x|x\in T\2\}\rightarrow \Complex$ with the usual scalar product.
$\|\pi_T(f)\|$ is the operator norm in that representation, which is bounded.
$C^*_{red}(\Gr)$
is called the algebra of the tiling $\tl$ and below denoted by $\A_\tl$.
It is separable, as $\Gr$ is second countable.
Note that the operator used in (\ref{14061}) belongs to such a representation,
in that $H=\pi_\tl(h)$ for some $h\in\A_\tl$.

\ms

Let us give a first rough characterization of $\A_\tl$
by the properties of the tiling. Remember that any $T\in\Om_\tl$ is
locally homomorphic to $\tl$, in the sense that
any \mi\ of $T$ does also occur in $\tl$, and that $T$ is locally isomorphic
to $\tl$
if the converse is true as well, i.e.\ if any pattern of $\tl$ occurs in $T$.
Moreover $\tl$ is homogeneous if it is locally isomorphic to any $T\in\Om_\tl$
which is equivalent to $\Om_T=\Om_\tl$ for all $T\in\Om_\tl$.
\bl
$\pi_T$ is faithful if and only if $T$ is locally isomorphic to $\tl$.
\el
{\em Proof:}
To prove faithfulness of $\pi_T$ we may restrict to elements of $C_c(\Gr)$.
Let $f\in C_c(\Gr)$.
If $\pi_T(f)=0$ then for all $\psi$ and $T'\sim T$:
$\sum_{T''}f(T',T'')\psi(T'')=0$. Hence $f$ vanishes on
$\{(T',T'')|T',T''\sim T\}$.
If $T$ is locally isomorphic to $\tl$ then $\{(T',T'')|T',T''\sim T\}$ is dense
in $\Gr$ so that $f$ has to vanish by continuity. If $T$ is not locally
isomorphic then $\tl$ contains a pattern $M$ which does not occur in $T$.
Let $\chi_{M,x}$ be the characteristic function on $U_{M,x}$ for some
$x\in M\2$, then $\pi_T(\chi_{M,x})=0$.\eb\bs

Since all the above representations are irreducible, and at least one of them,
namely $\pi_\tl$, is faithful, $\A_\tl$ cannot be semi-simple in case
$\tl$ is not homogeneous.
The ideal structure of $\A_\tl$ may be investigated
by studying the open invariant subsets of $\Om$, i.e.\ those open subsets
which contain next to an element $T$ all its equivalent elements. We will
just look at the simplest case in which $\Om_\tl$ does not contain
any proper invariant open subset
 which by \cite{Ren} implies simplicity of $\A_\tl$.
\bl
If $\tl$ is homogeneous then $\A_\tl$ is simple.
\el
{\em Proof:}
Let $U\subset\Om_\tl$ be open and invariant
hence $\Om_\tl\backslash U$ is invariant and closed.
If $U$ is not equal to $\Om_\tl$ let $T\in\Om_\tl\backslash U$. Then
$\Om_T\subset\Om_\tl\backslash U$ showing that
under the assumption of the lemma $U=\emptyset$.\eb\bs

Note that moreover for homogeneous
$\tl$ the algebra $\A_\tl$ is antilimial \cite{Ped}, namely
it is neither limial itself -- its primitive
spectrum contains one single point whereas the representations $\pi_T$ and
$\pi_{T'}$ are unitarily equivalent if and only if $T\sim T'$ -- nor can it
contain a limial ideal. It follows that the spectrum of any selfadjoint
$h\in\A_\tl$ has no discrete part, since the spectral projection onto
the eigenspace of a discrete eigenvalue would have to be represented by a
compact operator.\ms

Recall that any trace $\tr$ on $\A_\tl$ restricted to a linear functional
$\mu$ on $C(\Om)$ defines a
measure on $\Om_\tl$ also denoted by  $\mu$ through
$\mu(f)=\tr(f)=\int f d\mu$, which is invariant in the sense
that $\mu(U_{M,x})$ is independent of the \pkt\ $x$.
A direct consequence is that the inclusion
\be \label{11051}
\mu(C(\Om,\Z))\subset\tr_*K_0(\A_\tl)
\ee
is always valid.
Conversely any invariant normalized measure $\mu$ on the hull $\Om$
defines a normalized trace through
\be \label{24061}
\tr(f):=\int_\Omega\mbox{P}(f)\,d\mu
\ee
where $\mbox{P}:\,C^*_{red}(\Gr)\rightarrow C(\Omega)$ is the
restriction map. $\mbox{P}$ is the unique conditional expectation on
$C(\Om)$ and is faithful \cite{Ren}.
Moreover if $\tl$ is homogeneous every non trivial
invariant measure has to have closed support $\Omega$ so that $\tr$ defined
by (\ref{24061}) is faithful.
To obtain a gap labelling of a \So\ by means of the values
of its IDS a trace is required which satisfies
Shubin's formula. The question under which circumstances this is the case
for a given trace will not be addressed here, but see
 \cite{Be1,BBG,Ke2} for investigations in this directions.
In fact, conclusions on the nature of the spectrum may partly be drawn
without the need to connect the gap labelling with the values of
the IDS, any faithful trace may be used.
For instance if $\tr$ is faithful and the set of gap labels
${\cal L}_{\tr} (h)=\{\tr(\chi_{h\leq E})|E\notin\sigma(h)\}$ dense in $[0,1]$
the spectrum $\sigma(h)$ cannot contain a proper closed interval, for,
if $[a,b]\in\sigma(h)$, then by faithfulness
$\tr(\chi_{h\leq b})>\tr(\chi_{h\leq a})$ -- here $\tr$ has to be extended
to measurable functions over $\sigma(h)$ -- so that
$[0,1]\backslash{\cal L}_{\tr} (h)$
would contain the open interval $(\tr(\chi_{h\leq a}),\tr(\chi_{h\leq b}))$.
However, up to now there is no $K$-theoretic formulation of a condition for
a \So\ $h$
under which ${\cal L}_{\tr} (h)$ coincides with $\tr_*K_0(\A_\tl)$.

\section{Decorations of $\Z^d$}

\bd
A tiling shall be called a decoration of $\Z^d$ if there is a continuous
action $\varphi$ of $\Z^d$ on its hull $\Om$ such that $\Gr$ is isomorphic
as a topological groupoid to the transformation group $\Om\times_\varphi\Z^d$.
\ed
Isomorphic groupoids lead to isomorphic groupoid-\CA s.
To any continuous function
$f:\Om\times\Z^d\longrightarrow \Complex$ with compact support one may assign
the function $\hat{f}:\Z^d\longrightarrow C(\Omega)$ through
$\hat{f}(k)(T) = f(T,k)$. Carried over from   (\ref{25061},\ref{25062})
multiplication and involution then become convolution resp.\ involution twisted
by
$\varphi$:
\begin{eqnarray}
\hat{f}*\hat{g}\:(k)& =& \sum_{m\in\Z^d} \hat{f}(m)\,
(\hat{g}(k\!-\!m)\circ\varphi(m)) \\
\hat{f}^*\,(k)& =& \overline{\hat{f}(-k)\circ\varphi(k)}.
\end{eqnarray}
In fact, the closure
$C^*_{red}(\Om\times_\varphi\Z^d)$ is
isomorphic to $C(\Om)\times_\varphi\Z^d$,
the crossed product of $C(\Om)$ with $\Z^d$ by the action
$\varphi(k)(\hat{f}(m))=\hat{f}(m)\circ\varphi(k)$ \cite{Be1}.\ms

Decorations of $\Z^d$ do not have to be periodic. Any
one dimensional tiling is a decoration of $\Z$
as it may be understood as a two-sided sequence of (occasionally
decorated) intervals, i.e.\
as a map $\tl:\Z\rightarrow \{a_1,a_2,\cdots, a_n\}$, finiteness
of the alphabet $\{a_1,a_2,\cdots, a_n\}$ representing the intervals
being a consequence of the compactness condition.
$\varphi$ may be taken to be the left shift: $\varphi(T)_i=T_{i+1}$.
As well any $d$-fold Cartesian product of one dimensional tilings is a
decoration of $\Z^d$, but these are not the most general ones.\ms

The rest of this section is a longer remark which is not essential for the
sequel.\ms

Whether a tiling is a decoration of some $\Z^d$ or not seems in the examples
known to the author to be easy to check by verification of the requirements of
the definition.
However at some point one might wonder whether not all tilings
are $\Z$-decorations. In fact $\tl\2$ is countable, so why not
identifying it with $\Z$ or with $\Z^d$? The crucial point is, of course,
the topology. Let us make this point a little more precise.
Consider a bijection $\bet:\tl\2\rightarrow\Z^d$. Then
 $\beta:\Gamma\rightarrow\Z^d$ defines a surjective
homomorphism of groupoids through
$\beta(x,y)=\bet(y)-\bet(x)$,
and if we had neglected the closure in (\ref{21051}) $\Om$ would have to
be replaced by $\tl\2$ and
$\Gr$ by $\Gamma=\tl\2\times\tl\2$
which through $(\pi_1,\beta):\Gamma\rightarrow\tl\2\times_\varphi\Z$ is
 isomorphic to a
transformation group, the action being $\varphi(z)x=\bet^{-1}(\bet(x)-z)$.
The following theorem arose from the question under which circumstances
 $(\pi_1,\beta)$ extends to $\Gr$
giving rise to an isomorphism between $\Gr$ and $\Om\times_\varphi\Z^d$.
Let
\be
\Gamma(x,y):=\{(x',y')|M(x,y)=M(x',y')\},
\ee
$dist(x,y)$ be the diameter of $M(x,y)$, and $\pi_2$ denote projection onto
the second factor.

\bt
If a tiling $\tl$ allows for a bijection $\bet:\tl\2\rightarrow\Z^d$
which satisfies in addition
\begin{enumerate}
\item
for all $z\in \Z^d$ there is a constant $c_z$ such that $\forall
(x,y)\in\beta^{-1}(z): dist(x,y)<c_z$
\item
$\beta(x,y)=\beta(x',y')$ for all
$(x',y')\in\Gamma(x,y)$
\end{enumerate}
then it is a decoration of $\Z^d$.
\et
For the proof of the theorem we need two lemmas.
\bl       \label{20051}
If $\beta(x,y)=\beta(x',y')$ but $(x',y')\notin \Gamma(x,y)$ then
$U_{M(x,y),x}\cap U_{M(x',y'),x'}=\emptyset$.
\el
{\em Proof:}
Assume $\beta(x,y)=\beta(x',y')$ and
$T\in U_{M(x,y),x}\cap U_{M(x',y'),x'}$. Then there exists $x,y,v\in\tl\2$,
$v$ corresponding to the \pkt\ $y'$ once $x'$ has been identified with $0$
in $T$ such that $\beta(x,y)=\beta(x,v)$. Hence $y=v$ and therefore
$(x',y')\in \Gamma(x,y)$.\eb

\bl The map
$(\pi_1,\beta\circ\pi_2)\circ(\pi_1,\gamma)^{-1}$ is a homeomorphism between
$\Gr$
and $\Om\times \Z^d$.
\el
{\em Proof:}
First of all, since $\gamma(T,(x',y'))=\gamma(T,(x,y))$ for
$(x',y')\in\Gamma(x,y)$
\be \label{12052}
(\pi_1,\gamma)^{-1}(T,T-y)=\{(T,(x',y'))|(x',y')\in\Gamma(0,y)\}
\ee
which guarantees together with 2.\ that
$(\pi_1,\beta\circ\pi_2)\circ(\pi_1,\gamma)^{-1}$
is well defined, and Lemma~\ref{20051} shows that it is injective.
To show that it is bijective we construct a pre-image of $(T,z)$.
Any $T\in\Om$ is approximated by translates of $\tl$.
Choose $x$ such that $M_{c_z}(T)=M_{c_z}(\tl-x)$ and set
$y=\bet^{-1}(\bet(x)+z)\in\tl\2$. Then $(M(x,y),x)\subset(\tl-x,0)$
and 1.\ implies $(M(x,y),x)\subset(T,0)$ so that
$(\pi_1,\beta\circ\pi_2)(T,(x,y))=(T,z)$.

Moreover, because of
\be
(\pi_1,\beta\circ\pi_2)\circ(\pi_1,\gamma)^{-1}(U_{M,x,y})=U_{M,x}\times
\{\beta(x,y)\}
\ee
the map is open and we are left to show its continuity. Observe that by
Lemma~\ref{20051} and 1.\ the set
$S_{T'}(z)=\{(x,y)\in\beta^{-1}(z)|(M(x,y),x)\subset (T',0)\}$ is the same for
all $T'\in U_{M_{c_z}(T)}$. Therefore
\be
(\pi_1,\beta\circ\pi_2)^{-1}(U_{M_{c_z}(T)},z)
=U_{M_{c_z}(T)}\times S_{T}(z)
\ee
and, since the r.h.s.
is open, its image under $(\pi_1,\gamma)$ is open as well.\eb\bs

{\em Proof of the theorem:} The proof of the theorem is completed by
transforming the action of $\gamma$ into an action of $\Z^d$ on $\Om$.
For $(T,z)\in\Om\times\Z^d$ let $(x,y)\in S_T(z)$ and define
\be \label{13061}
\varphi(z)(T):=\gamma(T,(x,y))
\ee
which is again independent of the choice of $(x,y)$.
Continuity follows from the continuity of $\gamma(\cdot,(x,y))$.
To show that $\varphi(z_2)\circ\varphi(z_1)=\varphi(z_1+z_2)$ choose a
realization of $M_{c_{z_1}+c_{z_2}}(T)$ in $\tl$, let's say
$(\tilde{M},x)\subset(\tl\!-\!y,0)$ with
$M_{c_{z_1}+c_{z_2}}(T)=M_{c_{z_1}+c_{z_2}}(\tl\!-\!y)$.
Then $(\tilde{M},x)$ contains subpatterns
$M(x,x')$ and $M(x',x'')$ with $(x,x')\in S_T(z_1)$ and
$(x',x'')\in S_{T-(x'-x)}(z_2)$. Hence
$\varphi(z_2)\circ\varphi(z_1)(T)=\gamma(\gamma(T,(x,x')),(x',x''))
=\gamma(T,(x,x''))$ which is equal to $\varphi(z_1+z_2)(T)$ because of
$\beta(x,x'')=\beta(x,x')+\beta(x',x'')=z_1+z_2$. Thus $\varphi$ is an
action of $\Z^d$, and by (\ref{13061}) equivalence with respect to $\Gr$ may be
expressed as orbit equivalence under $\Z^d$.\eb

\section{Reductions of tilings and $K_0(\A_\tl)$}

In \cite{Ke2} all the \ti s of a tiling $\tl$ were considered to be \pkt d.
This was neccessary to obtain operators in $\A_\tl$ which
correspond to translations from a \ti\ to its
neighbours. But it turns out
to be quite useful to consider as well algebras which may be obtained
with fewer (or more) \pkt s. Restricting the allowed translations in a
controlled manner, resulting in what will be called a reduction
$\tl_r$ of $\tl$, furnishes us with an algebra $\A_{\tl_r}$ which is for
homogeneous $\tl$
\si\ to $\A_\tl$. The importance of this procedure lies in the possibility
that $\tl_r$ may be a decoration of $\Z^d$ which brings us a step closer to
the computation of $K_0(\A_\tl)$.\ms

Let $\tl_r\2$ be a subset of $\tl\2$ and set
$\Om_r=\overline{\{\tl-x|x\in\tl_r\2\}}$.
Let $\tl_r$ be the tiling which has
the same \ti s as $\tl$ but a different interpretation of the \pkt s.
The \pkt s of $\tl\2\backslash\tl_r\2$ shall either be interpreted as
decorations which might break symmetries or in case these symmetries are
not present just be neglected whereas the \pkt s of $\tl_r\2$ define the hull,
 i.e.\ $\Om_{\tl_r}=\Om_r$.
\bd
We call $\tl_r$ a reduction of
$\tl$ if $\Om_r$ is an open subset of $\Om_\tl$.
\ed
Recall that two \CA s $\A$ and $\B$ are called stably isomorphic if
$\A\otimes\K$ is isomorphic to $\B\otimes\K$, where $\K$ is the algebra of
compact operators (on a infinite dimensional separable Hilbert space).
\bl \label{22061}
Let $\tl_r$ be a reduction of a homogeneous $\tl$. Then $\A_{\tl_r}$
is \si\ to $\A_\tl$.
\el
{\em Proof:}
If $\tl_r$ is a reduction then $\chi_{\Om_r}\in\A_\tl$ and
$\A_{\tl_r}=\{a\in \A_\tl|a\chi^{}_{\Om_r}=\chi^{}_{\Om_r}a=a\}$.
Hence the latter is a hereditary subalgebra of the simple \CA\ $\A_\tl$.
By a theorem of \cite{Bro} hereditary subalgebras of simple separable
\CA s are \si.\eb\bs

Since $K_0(\A_\tl)$ is obtained via Grothendieck's construction
from the monoid of projector classes of $\A_\tl\otimes\K$,
the positive elements corresponding to the elements of the monoid,
it depends together with its order structure only on the stable
isomorphism class of $\A_\tl$. Moreover, in the above case the embedding
$\imath:\A_{\tl_r}\rightarrow\A_\tl$ induces an isomorphism
$\imath_*$ from $K_0(\A_{\tl_r})$ onto $K_0(\A_{\tl})$.
In fact, for separable \CA s, $\A$ being stably isomorphic to $\B$
is equivalent to the existence
of a (strong) Morita equivalence $\A$-$\B$-bimodule
which may be viewed as an element of $KK(\A,\B)$ and is a special
case of a $KK$-equivalence \cite{Bla,Ska,Cone}.
Any $KK$-equivalence between $\A$ and
$\B$ yields an isomophism of $KK(\Complex,\A)$ with $KK(\Complex,\B)$, namely
by multiplying it from the right, the multiplication being the Kasparov
product.
Translated into $K_0$-groups, $KK(\Complex,\A)$ being isomorphic to
$K_0(\A)$, the right multiplication of elements of $KK(\Complex,\A_{\tl_r})$
with the canonical Morita equivalence $\A_{\tl_r}$-$\A_\tl$-bimodule,
which as a linear space is
$\A_{\tl_r}\A_\tl$,
presicely  becomes $\imath_*$.

This shall now be used to extend the equality
\be \label{06061}
\mu(C(\Om,\Z))=\tr_*K_0(\A_\tl) ,
\ee
for any normalized trace $tr$ trace on $\A_\tl$ to a larger class of tilings
 including the Penrose and Ammann-Beenker tilings.
Equality (\ref{06061}) is so far
known for decorations of $\Z^d$ if $d\leq 3$ \cite{Els2}\footnote{
In \cite{Els2} ergodic measures have been used, but ergodicity not essential
for the proof of (\ref{06061}).}
and for arbitrary $d$ in case $\tl$ is a Cartesian product of
one dimensional tilings \cite{Ke2}.
Its proof is based on the identification $\A_\tl=C(\Om)\times_\varphi\Z^d$.
Since the crossed product by $\Z^d$ may be
understood as an iterated crossed product by $\Z$, the Pimsner Voiculescu exact
sequence \cite{PiVo} may be used for the computation of the $K$-groups
provided the sequence splits at a certain position, which
has been explicitly verified in \cite{Els2} up to $d=3$.
In case the tiling decomposes into a Cartesian product, the
groupoid decomposes, too, and $\A_\tl$ becomes a tensor product of
crossed products with $\Z$.
The K\"unneth formula then applies to reduce the $K$-groups to the
one dimensional case.
For higher $d$ the $K_0$-groups look more and more complicated but
their image under $\tr_*$ remains as simple as (\ref{06061}).
\bt \label{09061}
If a homogeneous tiling $\tl$ has a reduction $\tl_r$ which is a decoration of
$\Z^d$ with $d\leq 3$ or a Cartesian product of one dimensional tilings then
\be
\tr_*K_0(\A_\tl)=\mu(C(\Om,\Z))
\ee
where $\mu$ is the measure corresponding to the trace $tr$ on $\A_\tl$.
\et
{\em Proof:}
Denote by $\imath:\A_{\tl_r}\rightarrow\A_\tl$ the embedding. Then
$\tr\circ\imath$ is a trace
on $\A_{\tl_r}$ which is however not normalized,
$\tr(\imath(1_{\A_{\tl_r}}))=\tr(\chi^{}_{\Om_r})=\mu(\Om_r)$.
By Lemma~\ref{22061} $\A_\tl$ is \si\ to $\A_{\tl_r}$ and in particular
 $\imath_*K_0(\A_{\tl_r})=K_0(\A_{\tl})$.
But for $\tl_r$ which is a decoration of $\Z^d$ (\ref{06061})
can be used.
The invariant measure
on $\Om_r$ corresponding to $\tr\circ\imath$ being $\mu|_{\Om_r}$ this is
\be
\tr_*\circ \imath_*K_0(\A_{\tl_r})=\mu(C(\Om_r,\Z))
\ee
and therefore
\be
\tr_*K_0(\A_{\tl})=\mu(C(\Om_r,\Z))\subset \mu(C(\Om,\Z)).
\ee
Together with (\ref{11051}) the statement follows.\eb\bs

$K_0(\A_{\tl_r})$ and $K_0(\A_{\tl})$ differ only in their order units
(the images of the units of the algebras in $K_0$).
If one identifies them as above
the order unit of the former is $[\chi^{}_{\Om_r}]$.

\section{Applications: tilings obtained by the GDM}

Theorem~\ref{09061} has a wide range of applications. To show this we consider
tilings whose \ti s are $d$ dimensional
analogs of rhombi. We may or may not consider further
\mk ions of the rhombi e.g.\ to
break the symmetry. This is of importance for \sst\ tilings but is not
relevant below where we only show the existence of a reduction which is
a decoration of $\Z^d$.\bs

Given a set of $N$ (pairwise non parallel)
vectors $\alpha_1,\cdots,\alpha_N$ which span $\Real^d$ and a set
$J\subset \{1,\cdots,N\}$
containing $n$ elements,
an {\em $n$-facet of type $J$} is a subset of $\Real^d$ which is
translationally congruent to
$\{\sum_{i\in J}c_i\alpha_i|c_i\in [0,1]\}$.
We consider $n\leq d$ so these are $n$ dimensional
analogs of rhombi.
Let $\tl$ be a tiling of $\Real^d$ consisting
of \pe{d}s which are arranged
in such a way that \ti s touch, if they touch at all, at common complete
$d'$-facets, $d'<d$.
The \mii es of \ti s are thus the types of $d$-facets.
Let all \ti s of $\tl$ carry a \pkt.
If $d=N$ then $\tl$
itself is of course a decoration of $\Z^d$ so here we are
interested in $d<N$.
As before we assume that it satisfies the condition to ensure
compactness of its hull.

Now  consider the
following reduction of $\tl$.
Fix the type $I_0$ of the \ti\ which has \pkt\ $0$ and let
$\tl_r\2$ consist of all \ti s of this type. This is certainly a
reduction since $\Om_r=U_{a}$ with $a$ being of type $I_0$.
An {\em $i$-\faden}, $i\in I_0$, is an infinite line which is constructed as
follows: Join in any \ti\ which has \pe{d-1}s of type
${I_0\backslash\{i\}}$ their middle points by a line segment.
All these lines fit together to yield an infinite
line which we call an $i$-\faden, c.f.\ figure~\ref{10061}.
It is clear that never two \faden s of the
same type intersect.
Any type of \faden\ shall now be given a direction so that the \ti s
belonging to it can be ordered. Let us assume that
any $i$-\faden\ contains in both directions infinitely many \ti s of
type $I_0$.
Then
\be
\varphi_i(T)=T-x_i
\ee
defines a homeomorphism of $\Om_r$, where  $x_i$ denotes
the \pkt\ of the next \ti\ of type $I_0$
on the $i$-\faden\ to which $0$ belongs.
Clearly $\varphi$ is invertible. To proof its continuity denote by $V_i^{(n)}$
the set of all tilings such that the next \ti\ of type $I_0$
on \faden\ $i$ is the $n$'th one. By the above requirement $\Om_r=\bigcup_n
V_i^{(n)}$ and since $\Om_r$ is compact and
the $V_i^{(n)}$'s are open there is an $N$ such that
$\Om_r=\bigcup_{n\leq N}V_i^{(n)}$. In particular there is an upper bound $b_i$
for $dist(0,x_i)$ and
$\varphi_i(M_{r+2b_i}(T))\subset M_{r+b_i}(T-x_i)$.

For given $I_0$ there are $d$ types of \faden s furnishing
$d$ operations of $\Z$ on $\Om_r$. If
\bi
\item $\varphi_i\circ\varphi_j=\varphi_i\circ\varphi_j$ for all
$i,j\in I_0$
\ei
then $\varphi=\varphi_1\circ\cdots\circ\varphi_d$ is an operation of $\Z^d$ on
$\Om_r$ which is free and such that equivalence
with respect to the groupoid defined by $\tl_r$ becomes orbit equivalence, in
other words
$\tl_r$ is a decoration of $\Z^d$ (the topologies are easily seen to
coincide).\bs

A large class of tilings of the above kind which lead to
commuting actions may be constructed by the generalized
dual method (GDM) \cite{SoSt2} which we briefly explain.
A ($d$-dimensional, two-sided) linear {\em grid} (or more precisely $1$-grid)
is an infinite set of parallel
hyperplanes. They may be ordered involving a choice of direction and
two-sided refers to the requirement that
any hyperplane has a lower and an upper neighbour, i.e.\ the hyperplanes of
the grid may be given an integer number and any integer occurs.
A general grid is an infinite set of non-intersecting simply connected
unbounded hypersurfaces which may of course as well be ordered.
A linear $N$-grid consists of $N$ linear grids and each grid
(of type $i$) comes with
its so-called star vector $\alpha_i$, $i\in\{1,\cdots,N\}$,
which is normal to its
hyperplanes, of length one, and points towards the positive
direction of the order. They are all distinct and are supposed to
span $\Real^d$ ($N\geq d$).
More generally for arbitrary $N$-grids it is required that any $d$
hypersurfaces
of different grid type intersect at exactly one point. In this case choices
for the star vectors have to be made which have to be consistent \cite{SoSt2}.
An intersection point of $d$ hypersurfaces
(necessarily of different grid type)
is called regular if no other hypersurface intersects this point. It may
be as well classified by a certain type,
namely its type is $J$ which is
the collection of types of the grids to which the intersecting hypersurfaces
belong.

The actual tiling defined by such an $N$-grid is its dual in the following
sense.
Any intersection point (regular or not) corresponds to a \ti.
If the intersection point
is regular and of type $J$
then the \ti\ is a \pe{d} of type $J$ made from the star vectors.
The vertices correspond to the volumes surrounded by the hypersurfaces
and are given by
$\sum_{i=1}^N k_i\alpha_i$ where $k_i$ is the number of the lower
hypersurface of grid type $i$ surrounding the volume.

Now let us assume that
all intersection points of the chosen type $I_0$ are regular.
Then $\varphi_i$ may be carried over to an action on the $N$-grid.
Denoting by $g_i(k)$ the $k$'th hypersurface of grid $i$ the homeomorphism
$\varphi_i$
precisely becomes the shift from intersection point
$\bigcap_{j\in I}g_j(0)$ to intersection point
$\bigcap_{j\in I\backslash\{i\}}g_j(0)\cap g_i(1)$. Hence
$\varphi_i\circ\varphi_j$ is the shift from
$\bigcap_{k\in I}g_{k}(0)$ to
$\bigcap_{k\in I\backslash\{i,j\}}g_k(0)\cap g_i(1)\cap g_j(1)$ which is
certainly independent of the order. We thus have shown
\bt
Let $\tl$ be a $d$ dimensional
tiling which is obtained by the GDM from a regular
($d$ dimensional, two-sided, $N$-) grid.
Then it has a reduction which is a decoration of $\Z^d$.
\et

The introduction of decorations does not cause any problems. One may always
define $\tl_r\2$ to consist of those \ti s which would as undecorated \ti s
be of type $I_0$.\bs

The form of the Penrose tilings in which they appear as duals of linear
$5$-grids \cite{SoSt2} is slightly different from the form we used
in \cite{Ke2}. But this does not matter as the triangles used there always
form rhombi and choosing one of the triangle type (in its \mk ed version
where the mirror symmetry is broken)
amounts to the same as choosing a $2$-facet type.

The Ammann-Beenker tilings have as well a reduction which is a decoration of
$\Z^2$ as they are \mk ed variants of tilings obtained from  linear
$4$-grids. To give an explicit example
figure~\ref{10061} displays such a reduction where only those
squares carry a \pkt\ which have no horizontal link, i.e.\
$\tl_r\2$ contains precisely the squares in which
two ropes intersect.
For clarity this figure contains at the bottom the
prototiles, i.e.\ the congruence classes of the \ti s under all Euclidean
transformations (two of them are \mk ed).\bs

{\bf Acknowledgement.}
I thank Andreas H\"uffmann for very fruitful conversations.

\newpage

\listoffigures
\begin{figure}[p]
\unitlength1cm
\begin{picture}(0,1)
\end{picture}
\caption[Part of an Ammann-Beenker tiling with decoration and
\faden s. On the bottom the \mii es of the \ti s are shown.]{\label{10061}
Part of an Ammann-Beenker tiling with \faden s.}
\end{figure}

\end{document}